\documentclass[pra,amssymb,amsmath,twocolumn,superscriptaddress]{revtex4}
\usepackage{graphicx,amsbsy,amsmath,bbm}

\newcommand{\ket}[1]{\left| #1 \right\rangle}
\newcommand{\bra}[1]{\left\langle #1 \right|}
\newcommand{\bd}{{\bf d}}

\newcommand{\R}{{\mathbb{R}}}

\def\d{{\rm d}} 

\newcommand{\be}{\begin{equation}}
\newcommand{\ee}{\end{equation}}
\newcommand{\bea}{\begin{eqnarray}}
\newcommand{\eea}{\end{eqnarray}}
\newcommand{\beas}{\begin{eqnarray*}}
\newcommand{\eeas}{\end{eqnarray*}}
\newcommand{\refeq}[1]{Eq.~(\ref{#1})}

\newcommand{\ie}{{\it{i.e.}}}
\newcommand{\etal}{{\it{et al.}}}

\begin{document}
\title{Tests of multimode quantum non-locality with homodyne measurements}
\author{Antonio Ac\'\i n}
\affiliation{ICFO-Institut de Ciencies Fotoniques, Mediterranean
  Technology Park, 08860 Castelldefels (Barcelona), Spain }
\author{Nicolas J. Cerf}
\affiliation{QuIC, Ecole Polytechnique, CP 165, Universit\'e Libre de
Bruxelles, 1050 Brussels, Belgium}
\author{Alessandro Ferraro}
\affiliation{ICFO-Institut de Ciencies Fotoniques, Mediterranean
  Technology Park, 08860 Castelldefels (Barcelona), Spain }
\author{Julien Niset}
\affiliation{QuIC, Ecole Polytechnique, CP 165, Universit\'e Libre de
Bruxelles, 1050 Brussels, Belgium}
\begin{abstract}
  We investigate the violation of local realism in Bell tests involving
  homodyne measurements performed on multimode continuous-variable
  states. By binning the measurement outcomes in an appropriate way, 
  we prove that the Mermin-Klyshko inequality can be violated 
  by an amount that grows exponentially with the number of modes. 
  Furthermore, the maximum violation allowed by quantum mechanics 
  can be attained for any number of modes, albeit requiring a quantum state 
  that is rather unrealistic. Interestingly, this exponential increase of 
  the violation holds true even for simpler states, such as multipartite 
  GHZ states. The resulting benefit of using more modes is shown to be
  significant in practical multipartite Bell tests by analyzing 
  the increase of the robustness to noise with the number of modes. 
  In view of the high efficiency achievable with homodyne detection, 
  our results thus open a possible way to feasible loophole-free Bell
  tests that are robust to experimental imperfections. 
  We provide an explicit example of a three-mode state (a superposition
  of coherent states) which results in a significantly high violation of 
  the Mermin-Klyshko inequality (around $10\%$) with homodyne measurements.
\end{abstract}
\date{\today}
\maketitle
\section{Introduction}

The non-local nature of quantum mechanics is unarguably one of its most
counterintuitive aspects, marking a fundamental departure from the
classical picture of physical systems. First recognized by Einstein,
Podolsky, and Rosen \cite{EPR}, the intrinsic non-locality of the quantum
theory was put into a testable form via the celebrated Bell inequalities \cite{Bell}. Since the first experimental violation of a Bell inequality \cite{Aspect}, a great variety of tests have been performed. In all of them,
a local realistic model is shown to be incompatible with the actual 
experimental observations. Nonetheless, a fully conclusive evidence for rejecting local realism has not been achieved yet, for all the experiments
performed so far suffer from loopholes, forcing us to rely on
supplementary assumptions in order to reject local realism.
Specifically, a conclusive experiment should close two main loopholes: 
the locality loophole (namely, the measured
correlations must be collected with spacelike separated events)
and the detection-efficiency loophole (namely, the proportion between
detected and undetected events has to be high enough for the data
to be fully representative of the whole ensemble without the need to assume a 
``fair sampling'').

There is a large variety of quantum systems for which a test of local
realism may be envisaged. However, the quest for a loophole-free Bell test
has recently focused the research towards experiments involving
propagating light modes measured with homodyne detectors
\cite{Munro,Wenger,Patron,Nha}. The advantage of such 
a ``continuous-variable'' approach is twofold. 
First, light modes can be easily sent to spacelike separated
detectors suffering only a tolerable degree of decoherence. Second,
the current technology of homodyne detectors achieves a degree of
detection efficiency high enough to potentially close the detection
efficiency loophole. On the other hand, such an approach involves also
drawbacks whose resolution is still challenging. The main issue comes from the fact that the results of homodyne measurements can be described by means of the Wigner function. Thus, in order to avoid a local hidden variable description of the measured correlations, one should necessarily perform the test with a state endowed with a non-positive Wigner function (otherwise, the Wigner function
is a genuine probability distribution, which provides an explicit 
local realistic model of the data). Even if single-mode states of traveling light have already been generated with a non-positive Wigner function 
\cite{NGaus_exp1,NGaus_exp2}, such (non-Gaussian) states are hard to master in the laboratory. Among the above mentioned
proposals of Bell tests with homodyne detection, those nearer 
to an experimental realization only give a small violation of Bell 
inequalities \cite{Patron,Nha}, whereas higher violations 
involve states whose actual generation seems rather unrealistic \cite{Wenger}. In this context, the search for Bell tests involving
feasible resources and giving, at the same time, violations that are high
enough to be robust against experimental noise is very desirable.
This is the thrust of the present paper.

Bell tests relying on homodyne detection typically involve the
following scenario.  Two parties perform
spacelike separated homodyne measurements 
by randomly choosing each between two settings, thus measuring two quadratures of the incoming electromagnetic fields. The violation of the Bell Clauser-Horne-Shimony-Holt (CHSH) inequality \cite{CHSH} is then tested.
Since this inequality is devised for two-outcome
measurements, the collected data, which are distributed (ideally) in a
continuous way, have to be discretized; such a
procedure is referred to as binning process.  In this work, we will
generalize this scenario to more than two parties, considering the
Mermin-Klyshko (MK) type of Bell inequalities, which involves $m$ parties,
two measurement settings, and two outcomes \cite{Mermin_Klyshko}. Our
main motivation is that higher violations of local realism are
expected as the number of parties increases \cite{GBP98}, which may
help in the search for feasible proposals of loophole-free Bell tests
able to tolerate the experimental noise.

The earlier investigations of Bell tests in infinite-dimensional spaces
with multipartite settings have considered several scenarios. In
Ref.~\cite{VLoock}, a test based on the measurement of the light field
parity was envisaged. It was found that the violation does not increase
exponentially, as one would have expected, but this may be due to the
fact that no optimization over the possible measurement settings was
performed \cite{job}. In Ref.~\cite{pseudospin} instead, a maximal
violation of Mermin-Klyshko inequality was found for continuous
variable states, considering the measurement of a different class of
operators which can be seen as the continuous-variable analogue of the
spin operators. However, both these approaches deal with non-Gaussian
measurements described by a non-positive Wigner function, which are
far from the reach of current detection technology. Recently,
another approach has been introduced that does not
rely on the use of inequalities with discrete outcomes, thus avoiding
the need of a binning procedure \cite{Cavalcanti}.  
There, quadrature measurements via
homodyne detection are considered, and the measurement outcomes are
used to directly test a novel Bell inequality for continuous
variables. Interestingly, the authors find that the violation of the
local realistic bound is exponential in the number of parties.
However, a possible experimental implementation is still very
challenging as it would require at least ten space-like separated
homodyne measurements. It is then unclear whether such a novel approach can
give advantages, from a practical perspective, over non-locality tests
involving binning strategies.

In this paper, we will stick with the use of discrete-variable Bell
inequalities in the tests of non-locality for continuous-variable states, 
thus using some binning procedure. 
We will address the following question: {\it Is it possible
to have an exponential increase of the violation of local realism in a
test involving quadrature measurements of $m$ modes and considering
the Mermin-Klyshko inequalities? }
We will answer this question by the affirmative, providing
specific examples of states exhibiting such a behavior. 
These states belong to the class of photon-number correlated states,
and generalize to an arbitrary number of modes the approach of Ref.~\cite{Munro}. Remarkably, we will show that it is even possible to reach the maximal violation of the Mermin-Klyshko inequalities for $m$ modes
by properly choosing the binning procedure and the quantum state, thereby extending the results obtained in Ref.~\cite{Wenger} for two modes.

The theoretical interest of our results resides in the
fact that we are not considering a direct mapping 
from continuous- to discrete-variable states. Thus, the possibility to
obtain a maximal violation of Mermin-Klyshko inequality for any number
of parties is {\it a priori} non-trivial. 
Furthermore, we may have anticipated that the binning procedure causes 
an irreparable loss of information, preventing the possibility to
reach a maximum violation. 
From an experimental perspective, our results imply that various strategies
allow for an increase of the violation of locality in homodyne-based 
Bell tests. This, in turn, gives specific insights in the search of multipartite states appropriate for an actual experiment.

The paper is organized as follows. In Sec.~\ref{cpns}, we analyze
a Bell test involving multimode states with perfect correlations in
the number of photons in all modes. 
We analytically show that the violation 
of local realism can grow exponentially with the number of
modes involved.  This is true despite the fact that a simple sign
binning strategy is used. Nonetheless, we see that the resulting violation 
does not reach the maximum value achievable within quantum mechanics. 
However, we show in Sec.~\ref{mv} that a different binning strategy, 
properly tailored for
some specific states, allows attaining the maximal violation of
locality achievable within quantum mechanics. Then, in
Sec.~\ref{noise}, we address the robustness of these
tests of nonlocality. 
We see that, at least for the analyzed noise model, the tolerable noise
increases with the number of modes, making the violation 
better testable in practice.
In Sec.~\ref{tmsg}, we exploit all these results from a
practical oriented perspective. In particular, we give an
explicit example of a class of three-mode states which, on the one hand, exhibit
a gain in the violation by considering more than two
modes, and, on the other hand, may be implemented with the near future
technology. Finally, we close the paper in Sec.\ref{esco} with
some concluding remarks.

\section{Correlated photon number states}\label{cpns}

We show in this Section that in order to obtain an exponential
violation of local realism with quadrature measurements, it is
possible to rely on a simple binning strategy and the standard
Mermin-Klyshko Bell inequalities. In achieving this goal we generalize
to $m$ parties the work of Ref.~\cite{Munro}, where the bipartite
scenario was analyzed.

Before proceeding, let us recall the general form of the Mermin-Klyshko
Bell inequalities. Consider two dichotomic observables
$O_t$ and $O_t'$ for each party $t$.
The Mermin-Klyshko inequalities are based on the
recursive definition of the Bell operator
\begin{align}
B_t \equiv
 \frac {B_{t-1}}{2} \otimes ( O_t+ O_t' ) 
 +\frac {B'_{t-1}}{2} \otimes ( O_t- O_t' ) ,
\label{bell_op}
\end{align}
where $B_1=2\, O_1$, $B_1'=2\, O_1'$, and $B'_t$ denotes the
same expression as $B_t$ but with all $O_t$'s and $O_t'$'s
exchanged \cite{GBP98}. The Mermin-Klyshko inequality for $m$
parties then reads
\begin{align}
{\cal  B}_m\equiv|\langle B_m \rangle|\le 2\,.
\label{mk}
\end{align}
In the case of two parties ($m=2$), this expression reduces to 
the well-known CHSH inequality, for which
\begin{align}
B_2 =  O_1\otimes O_2+ O_1\otimes O_2'+ O_1'\otimes O_2 -O_1'\otimes O_2'
\end{align}

Let us now consider a generic photon-number correlated state of $m$
bosonic modes
\begin{equation}\label{munro_state}
  \ket{\Psi}=\sum_{r=0}^\infty c_r \ket{r}_1\ket{r}_2...\ket{r}_m \,,
\end{equation}
with $\sum_r |c_r|^2=1$. 
The two observables $O_t$ and $O_t'$ to be measured are chosen as
the two quadratures of the electromagnetic field $X(\theta_t)$ 
and $X(\theta_t')$ (corresponding to angles $\theta$ and $\theta'$).
These can be measured by applying an homodyne detection
on each mode $t$. In the following, we use 
the notation $X_t$ for the quadrature $X(\theta_t)$,
and $x_t$ for the outcome of its measurement (and similarly
for primed quantities).

The joint probability to obtain the outcomes $x_1,...,x_m$ 
when measuring the quadratures $X_1,...,X_m$ is given by
\begin{align}\label{jointp}
  {\cal P}(x_1,...,x_m) =
& \left|
_1\!\bra{x_1}...\,_m\!\bra{x_m}\Psi\rangle
\right|^2 =
 \sum_{r,s=0}^\infty c_r\,c^*_s \nonumber \\
& \times \frac{e^{i\phi(r-s)}}{(\pi 2^{r+s}r!s!)^{m/2}}
\prod_{t=1}^{m}e^{-x_t^2}H_r(x_t)H_s(x_t)\;,
\end{align}
where $\phi=\theta_1+\cdots \theta_m$, $H_t(x)$ is the Hermite polynomial 
of degree $t$, and $\ket{x_t}$ are the eigenvectors of
the quadrature operator $X_t$.

Consider the following simple binning strategy, known as {\it sign binning}:
when the result of a quadrature measurement falls in the domain
$\mathbb{R}^\pm$ the value $\pm1$ is associated to it.  Now, let us
start by calculating the probability ${\cal P}_{+1,...,+1}$ that a ``$+1$''
result is observed in all the measuring sites:
\begin{eqnarray}\label{allplus}
{\cal P}_{+1,...,+1} &=& \int_{0}^{\infty}\!\,\d x_1 \cdots \int_{0}^{\infty}\!\,\d x_m {\cal P}(x_1,...,x_m)
\nonumber \\
&=& \sum_{r,s=0}^\infty c_r\,c^*_s \frac{e^{i\phi(r-s)}}{(\pi 2^{r+s}r!s!)^{m/2}}
\nonumber \\
& & \times \prod_{t=1}^{m}\int_{0}^{\infty}\!\,\d x_t\, e^{-x_t^2}H_r(x_t)H_s(x_t)\,.
\end{eqnarray}
The integrals above can be evaluated by recalling the following
properties of Hermite polynomials for $r\neq s$,
\begin{equation}
\int_{0}^{\infty}\!\,\d x \, e^{-x^2}H_r(x)H_s(x) = \frac{\pi 2^{r+s}}{r-s}[F(r,s)-F(s,r)] \,,
\end{equation}
where we defined $F(r,s)$ as
\begin{equation} F(r,s)^{-1}=\Gamma\left(\frac{1}{2}-\frac{1}{2}r\right)\Gamma\left(-\frac{1}{2}s \right)\,,
\end{equation}
with $\Gamma$ being the Gamma function. For $r=s$, one has instead
\begin{equation}
  \int_{0}^{\infty}\!\,\d x \, e^{-x^2}H_r^2(x) = 2^{r-1} \, r!\, \sqrt\pi \,,
\end{equation}
Defining the functions
\begin{eqnarray}\label{gfunction}
{\cal G}(\phi,m)&=&2 \sum_{r>s} \mathrm{Re}(c_r c^*_s)\, g_{r,s}(\phi,m)
 \,, \nonumber \\
g_{r,s}(\phi,m)&=& \left(\frac{\pi\, 2^{r+s}}{r!s!}\right)^{m/2}\left[\frac{F(r,s)-F(s,r)}{r-s}\right]^m 
 \nonumber \\
& & \times \cos[\phi(r-s)]\,,
\end{eqnarray}
one obtains that
\begin{equation}
{\cal P}_{+1,...,+1}=\frac{1}{2^m}+{\cal G}(\phi,m)\,.
\end{equation}
The other probabilities can be obtained in a similar way. Let us
define the multi-index $\bd=(d_1,\cdots d_m)$, with $d_t=\pm 1$ denoting
the measurement outcome obtained for mode $t$ after binning. Then,
the joint probability for a generic collection $\bd$ of measurement outcomes
will be indicated by ${\cal P}_\bd$. It can be obtained 
similarly as ${\cal P}_{+1,...,+1}$ by recalling
that an Hermite polynomial of even (odd) degree is an even (odd)
function, namely
\begin{equation}\label{Pd}
  {\cal P}_\bd=\frac{1}{2^m}+\sigma(\bd) \, {\cal G}(\phi,m)\,,
\end{equation}
where $\sigma(\bd)=\prod_{t=1}^{m}d_t$. Now, we are in the position to
calculate the generic correlation function between the measurement
results $E(\phi,m)$ (note that it only depends on the sum
of the angles $\phi$). By definition, we have
\begin{equation}\label{Edef}
  E(\phi,m)=\sum_{\bd}\sigma(\bd) \, {\cal P}_\bd\,,
\end{equation}
where the sum goes over all possible collections $\bd$
of measurement outcomes. Since
the number of possible collections for which $\sigma(\bd)=1$ is equal to
that for which $\sigma(\bd)=-1$, one finally has, by substituting \refeq{Pd}
into \refeq{Edef}, that
\begin{equation}\label{EofG}
  E(\phi,m)=2^m \, {\cal G}(\phi,m)\,.
\end{equation}

Let us show that we can reach an exponential violation with a simple analytically manageable example. For the simple case of three parties ($m=3$), the Mermin-Klyshko inequality then reads
${\cal  B}_3\equiv|\langle B_3 \rangle|\le 2$, with
\begin{align}
B_3= & O_1 \otimes O_2 \otimes O_3'
+O_1 \otimes O_2' \otimes O_3
+O_1' \otimes O_2 \otimes O_3 \nonumber \\
& -O_1' \otimes O_2' \otimes O_3'
\label{MK3}
\end{align}
Considering a tripartite GHZ state,
\begin{equation}\label{ghz_3}
  \ket{{\rm GHZ}_3}=\frac{1}{\sqrt2}(\ket{000}+\ket{111}) \, ,
\end{equation}
we have $c_0=c_1=2^{-1/2}$ and $c_{r\ge 2}=0$, so that
${\cal G}(\phi,3)=g_{1,0}(\phi,3)=(2\pi)^{-3/2} \cos(\phi)$
and \refeq{EofG} becomes
\begin{equation}
E(\phi,3)=2^3 {\cal G}(\phi,3) = (2/\pi)^{3/2} \cos(\phi).
\end{equation}
The GHZ-like angles ($\theta_1=0$, $\theta_2=\pi/6$, $\theta_3=2\pi/6$, and $\theta_i'=\theta_i+\pi/2$), give the maximum violation of the inequality, namely 
\begin{equation}
{\cal  B}_3=|3\, E(\pi,3)-E(0,3)|=4 \left(\frac{2}{\pi}\right)^{3/2}\simeq 2.032 \, . 
\end{equation}
Now, consider the multipartite generalization of the GHZ state
\begin{equation}\label{mGHZ}
  \ket{{\rm GHZ}_m}=\frac{1}{\sqrt2}(\ket{0...0}+\ket{1...1}) \,,
\end{equation}
\refeq{EofG} becomes
\begin{equation}\label{EmGHZ}
  E(\phi,m)=\left(2/\pi\right)^{m/2}\cos\phi \,.
\end{equation}
The dependence on the angle $\phi$ of the above correlations is the
same as the one appearing in a standard spin-like 
test for a multipartite GHZ state, namely $E(\phi,m)=\cos(\phi)$. 
In that case, it is known that the choice of GHZ-like angles 
$\theta_k=(-1)^{m+1}\pi(k-1)/(2m)$ and $\theta_k'=\theta_k+\pi/2$,
gives the highest value of the Bell factor, namely $2^{(m+1)/2}$ (see e.g.
Ref.~\cite{GBP98}). Therefore, using the same angles,
the corresponding Bell factor reads
\begin{equation}\label{BmGHZ}
  {\cal B}_m=\sqrt2 \, \left(4/\pi\right)^{m/2} \,,
\end{equation}
giving rise to an exponential violation of local realism.\\

Apart from this simple analytical example, one can use a numerical approach to show the exponential violation of local realism as formula (\ref{EofG}) is easily amenable to perform numerical calculations for a fixed number of parties $m$. 
In order to find the state $\ket{\Psi}$ (coefficients $c_r$'s) that maximally violates the Mermin-Klyshko inequality,
one has to evaluate the corresponding Bell factor $B_m$ 
for a given configuration of measuring angles. The Bell factors 
are expressed in general by a linear combination of correlation functions 
given each by \refeq{EofG}, with the prescription given in \refeq{bell_op}. 

\begin{figure}
  {\includegraphics[width=0.45\textwidth,height=0.3\textwidth]{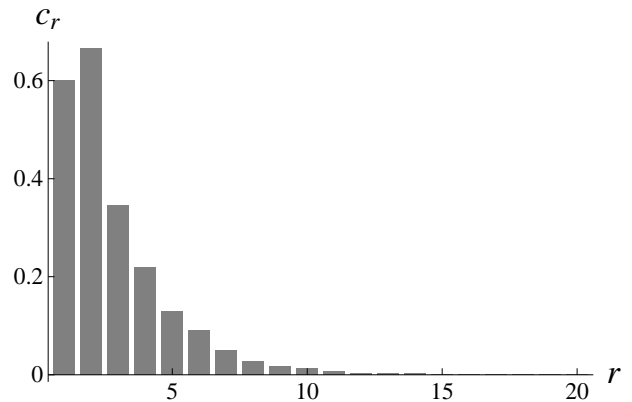}}
  \caption{Coefficients $c_r$ for the optimal state $\ket{\Psi}$ in the case
    of $m=3$, $d=20$ (with GHZ-like measurement angles). The Bell
    factor is ${\cal B}_3\simeq2.204$.}
\label{B3optpsi}
\end{figure}

Let us search the state which maximizes 
the violation of the Mermin-Klyshko inequality for the particular GHZ-like choice of angles.
For three modes ($m=3$), defining the (infinite dimensional) real 
symmetric matrix $B_3$ as
\begin{equation}\label{B3ex}
  [B_3]_{r,s}=2^3 \, (3g_{r,s}(\pi,3)-g_{r,s}(0,3))\,,
\end{equation}
with the diagonal elements being set to zero,
we note that the Bell factor can be re-expressed 
as ${\cal B}_3=C^\dag B_3C$, where the elements of the vector $C$ 
are given by the coefficients of the input state, \ie~$[C]_r=c_r$. 
Consequently, 
the maximal violation of the Mermin-Klyshko inequality is simply given 
by the maximal eigenvalue of the matrix $B_3$, while the optimal input state 
is determined by its corresponding eigenvector. In order to perform a numerical
analysis, one has to truncate the Hilbert space dimension of
$\ket{\Psi}$ to some arbitrary $d$. For example, for $d=2$ the optimal
choice turns out to be the GHZ state (\ref{ghz_3}),
giving a violation of ${\cal B}_3=2.032$. By increasing the dimension $d$,
the asymptotic violation is given by ${\cal B}_3\simeq 2.205$. In
Fig.~\ref{B3optpsi}, we show the coefficients $c_r$ for the optimal
state $\ket{\Psi}$ in the case $d=20$, for which the Bell factor is ${\cal
  B}_3\simeq2.204$.

\begin{figure}
  {\includegraphics[width=0.45\textwidth,height=0.3\textwidth]{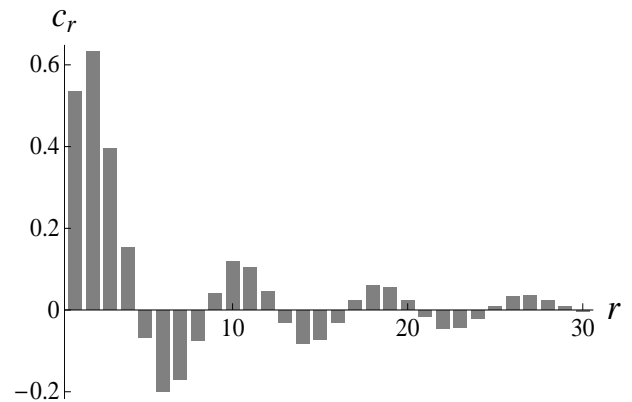}}
  \caption{Coefficients $c_r$ for the optimal state $\ket{\Psi}$ in the case
    of $m=2$, $d=30$. The Bell factor is ${\cal B}_2\simeq2.1$.}
\label{B2optpsi}
\end{figure}

The same procedure can be applied for any number of parties.
In the case of two parties ($m=2$), one recovers the results given by Munro in
Ref.~\cite{Munro} provided that the constraint $c_r>0$ is taken into account,
namely ${\cal B}_2\simeq 2.076$.
Interestingly, a higher violation can be achieved if we
consider negative coefficients for $\ket{\Psi}$. As an example, we report in
Fig.~\ref{B2optpsi} the coefficients $c_r$ for the optimal state in
the case $d=30$, for which the Bell factor raises up to ${\cal B}_2\simeq 2.100$
(recall that the Bell factor can be written in this case as ${\cal
  B}_2=3E(\phi,2)-E(3\phi,2)$, where we have chosen
  $\phi=\pi/4$, as in \cite{Munro}).

\section{Maximal violation}\label{mv}

Let us now see if we can find a class of states and a binning strategy
that allows for a maximal violation of the Mermin-Klyshko inequality
for any number of parties $m$ and using quadrature measurements. We recall that the maximal quantum violation of these inequalities is given by ${\cal B}_m^{max}=2^{(m+1)/2}$ \cite{GBP98}. Inspired by
the results of Ref.~\cite{Wenger}, we introduce the state
\begin{align}\label{mv_state}
  \ket{\Psi}=\frac{1}{\sqrt{2}}(|f\rangle^{\otimes m} +
  e^{i\theta}|g\rangle^{\otimes m})\;,
\end{align}
where $f$ is a real and even function of some quadrature $x$, while $g$ is real and odd (thus, $f$ and $g$ are orthogonal). Both functions are also normalized 
to unity. Note that because $f(x)$ is real and even, it has a real and even Fourier transform $\tilde{f}(p)$, while $g(x)$ has an imaginary Fourier transform $i\tilde{h}(p)$, with $\tilde{h}(p)$ real and odd.

Each party $t$ chooses to measure one of two conjugated quadrature via homodyne detection, either $X(0)=X$ or $X(\pi/2)=P$, and obtains a continuous variable $x_t$ or $p_t$ depending on the choice of measurement setting. Suppose the $k$ first parties measure the $X$ quadrature, while the remaining $m-k$ measure the conjugate quadrature $P$. The joint probability that they obtain the results $x_1,...,x_k,p_{k+1},...,p_m$ is given by
\begin{align}\label{prob}
\nonumber {\cal P}(x_1,...&,x_k,p_{k+1},...,p_m)=|\langle x_1|...\langle x_k|\langle p_{k+1}|...\langle p_m |\Psi\rangle |^2\\
\nonumber = &\frac{1}{2}\big( f^2(x_1)...f^2(x_k)\tilde{f}^2(p_{k+1})...\tilde{f}^2(p_m)\\
\nonumber&+g^2(x_1)...g^2(x_k)\tilde{h}^2(p_{k+1})...\tilde{h}^2(p_m)\\
\nonumber&+2\cos(\theta+(m-k)\frac{\pi}{2})
\times f(x_1)g(x_1)...f(x_k)g(x_k)\\
&\times \tilde{f}(p_{k+1})\tilde{h}(p_{k+1})...\tilde{f}(p_m)\tilde{h}(p_m)\big)\;.
\end{align}
To exploit the parity properties of $f$ and $g$, we introduce the root binning defined in Ref.~\cite{Wenger}. This binning depends on the roots of the known functions $f$ and $g$. If party $t$ measures the $X$ quadrature, the result will be interpreted as a ``$+1$'' if the measured value $x_t$ lies in the interval where $f(x_t)$ and $g(x_t)$ have the same sign, and ``$-1$'' if their signs are opposite, i.e. we consider the following domains
\begin{align}
\nonumber D_{x}^+&=\{x\in\R|f(x)g(x)\geq 0\}\\
D_{x}^-&=\{x\in\R|f(x)g(x)< 0\}\;.
\end{align}
We can similarly define the domains $D_{p}^+$ and $D_{p}^-$ associated to the measurement of the quadrature  $P$.
For the choice of measurement settings defined above, we can thus calculate $2^m$ probability ${\cal P}_\bd$ corresponding to the observation of a given collection  $\bd$ of binary results. For example, the probability ${\cal P}_{+1,...,+1}$ that each party observes a ``$+1$'' result reads
\begin{align}
\label{pquad}
\nonumber {\cal P}_{+1,...,+1}=\int_{D_{x}^+}&\d x_1 ... \int_{D_{x}^+}\d x_k\int_{D_{p}^+}\d p_{k+1}...\int_{D_{p}^+}\d p_m\\
 &\times {\cal P}(x_1,...,x_k,p_{k+1},...,p_m)\;.
\end{align}
We are now in the position to calculate the correlation function $E(X_1,\cdots X_k,P_{k+1},\cdots P_m)$. Note that since $f$ and $g$ ($\tilde{f}$ and $\tilde{h}$) are even and odd respectively, $f^2$ and $g^2$ ($\tilde{f}^2$ and $\tilde{h}^2$) are even functions. Hence the first two terms of the right hand side of (\ref{prob}) are even functions also, and their contribution to the correlation function will vanish. We thus obtain the remarkably simple expression
\begin{align}\label{corr}
\nonumber E(X_1,\cdots X_k,P_{k+1}, \cdots P_m)&=V^k W^{m-k} \\
&\times \cos\left[\theta + (m-k)\frac{\pi}{2}\right]
\end{align}
where
\begin{align}\label{VW}
\nonumber V &=\int_{-\infty}^\infty \d x|f(x)g(x)|\\
W &=\int_{-\infty}^\infty \d p|\tilde{f}(p)\tilde{h}(p)|
\end{align}
Interestingly, the correlation function (\ref{corr}) only depends on the number of sites where $X$ and $P$ are measured. All correlation functions corresponding to $k$ measurements of the $X$ quadrature, and $m-k$ measurements of the $P$ quadrature are therefore equal. We will denote them $E(k,m-k)$ to emphasize this property.

Let us illustrate the power of this compact notation with an example. For $m=3$, the Bell factor reads
\begin{align}\label{B3}
\nonumber {\cal  B}_3 =&|E(X_1,X_2,P_3) + E(P_1,X_2,X_3) + E(X_1,P_2,X_3)\\
\nonumber &-E(P_1,P_2,P_3)|\\
\nonumber =&|3 E(2,1) - E(0,3)|\\
=&|3 V^2 W \cos(\theta + \frac{\pi}{2}) - W^3 \cos (\theta +3\frac{\pi}{2} )|
\end{align}
We see that the maximal violation, i.e. ${\cal  B}^{max}_3 = 4$, can be reached with a state $|\Psi\rangle$ such that $\sin(\theta)=\pm1$ and $V=W=1$. Although such a state is quite unrealistic, one can define a family of physical states that approximates it arbitrarily well. The corresponding $f$ and $g$ functions are trains of gaussians, and $V,W\rightarrow 1$ as the number of peaks goes to infinity. We refer the reader to Ref.~\cite{Wenger} for their exact analytical expression.

Let us now generalize this result for an arbitrary $m$. First note that the Bell factor (\ref{mk}) can be written as
\begin{align}\label{bellf}
\nonumber{\cal B}_m = \frac{1}{2}|\langle X_{m} B_{m-1}\rangle + \langle P_{m} B_{m-1}\rangle &+\langle X_{m} B'_{m-1}\rangle\\
&- \langle P_{m} B'_{m-1}\rangle |
\end{align}
with $B_1=2\, X_1$ and $B_1'=2\, P_1$.
In order to benefit from our compact notation, we explicitely expand 
the expectation values of $B_{m-1}$ and $B'_{m-1}$ in terms of correlation functions
\begin{align}\label{B}
\nonumber \langle B_{m-1}\rangle &= \sum_{k=0}^{m-1} \alpha_k E(k,m-1-k)\\
 \langle B'_{m-1}\rangle &= \sum_{k=0}^{m-1} \alpha_k E(m-1-k,k)
\end{align}
where the $\alpha_k$'s are some known coefficients. When $m-1=3$ for example, we have $\alpha_1=3$, $\alpha_3=-1$, and $\alpha_0=\alpha_2=0$.
As the correlation function only depends on the number of $X$ and $P$ measurements, the average values of the four operators of (\ref{bellf}) can be easily calculated from $\langle B_{m-1}\rangle $ and $\langle B'_{m-1}\rangle $.
Suppose $B_{m-1}$ has a term proportional to the $X_1 ...X_kP_{k+1}... P_{m-1}$ operator, which leads to the correlation function $E(k,m-1-k)$. The operator $X_mB_{m-1}$ will thus have a term proportional to $X_m X_1 ...X_kP_{k+1}... P_{m-1}$ leading to the correlation function $E(k+1,m-1-k)$, i.e. at the level of corelation functions we only need to replace $k$ by $k+1$ as the $X$ quadrature is measured at one additional site.
A similar argument for the expectation values of $P_mB_{m-1}$, $X_mB'_{m-1}$ and  $P_mB'_{m-1}$ leads to
\begin{align}\label{bellfcc}
\nonumber {\cal B}_m =  & \frac{1}{2}\Big| \sum_{k=0}^{m-1} \alpha_k [ E(k+1,m-1-k)+E(k,m-k) \\ 
+&E(m-k,k)-E(m-k-1,k+1) ] \Big|
\end{align}
To maximize this expression, we note that for two and three parties the maximum violation is reached for a state with $V=W=1$. We thus make the reasonable hypothesis that it remains true for an arbitrary $m$. Recall that we know how to choose $f$ and $g$ such as to reach these values. When $V=W=1$, we have
\begin{align} \label{Ekmk}
E(k,m-k) &= \cos\left[\theta + (m-k)\frac{\pi}{2}\right]
\end{align}
and \refeq{B} becomes
\begin{align}
\label{BVW1}\langle B_{m-1}\rangle &= \sum_{k=0}^{m-1} \alpha_k \cos\left[\theta + (m-1-k)\frac{\pi}{2}\right] \nonumber \\
\langle B'_{m-1}\rangle &= \sum_{k=0}^{m-1} \alpha_k \cos\left[\theta+k\frac{\pi}{2}\right]
\end{align}
Introducing Equation~(\ref{Ekmk}) into Eq.~(\ref{bellfcc}) and using some well known trigonometric formulae, the Bell factor simplifies to
\begin{align}
{\cal B}_m=& \left|\cos\left(\theta + m\frac{\pi}{4}\right) + \sin\left(\theta + m\frac{\pi}{4}\right) \right|
\nonumber\\
& \times \left| \sum_{k=0}^{m-1}\alpha_k \cos\left[(m-2k)\frac{\pi}{4}\right] \right|
\end{align}
Maximizing the violation of local realism boils down to finding the optimal phase $\theta_m$ such that the first factor of the right hand side is maximum. This term achieves its maximum of $\sqrt{2}$ for a value of the phase
\begin{equation}\label{theta}
 \theta_m=(1-m)\frac{\pi}{4}
\end{equation}
We also note that
\begin{align}
 (m-2k)\frac{\pi}{4} = \theta_{m-1} + (m-1-k)\frac{\pi}{2}
\end{align}
hence the maximum value of the Bell factor can be finally written as
\begin{align}
\label{eq:33} {\cal B}_m^{max}&=\sqrt{2} \, \left|\sum_{k=0}^{m-1}\alpha_k \cos \left[\theta_{m-1} + (m-1-k)\frac{\pi}{2}\right]\right|\\
&= \sqrt{2}\hspace{2mm} {\cal B}_{m-1}^{max}
\end{align}
where we have identified the summation of Eq.~(\ref{eq:33}) with Eq.~(\ref{BVW1}) at the optimal angle $\theta_{m-1}$. Introducing now the maximum value obtained in Ref.~\cite{Wenger} for the two party case, ${\cal B}_2^{max}=2\sqrt{2}$, we obtain by recursion
\begin{align}
{\cal B}_m^{max}=2^{(m+1)/2}
\end{align}
which is the known maximal bound imposed by quantum mechanics. Remarkably, the state $|\Psi\rangle$ defined in (\ref{mv_state}) combined with homodyne detection and a binning strategy called \textit{root binning} allows for a maximal violation of the MK inequality. This result shows that even if the binning process discretizing the result of the homodyne detection discards some information, it does not prevent to maximally violate tests of local realism based on discrete variables.

\section{Effect of noise}\label{noise}

In Sections \ref{cpns} and \ref{mv}, we proved that the search of loophole-free Bell tests might benefit from an increased number of parties involved in the experiment. The signature of this improvement lies in the exponential increase of the Bell factor with the number of parties $m$. However, what makes a Bell test challenging in practice is not the magnitude of the violation, but rather the inevitable noise associated with any real experiment. In many cases, these imperfections are sufficient to hide the non-local correlations that one tries to observe. When the number of parties involved in a Bell test increases, so does the fragility of the state used in the experiment. The risk is thus to rescale the violation in such a way that no benefit of a larger $m$ is witnessed in practice. One can therefore correctly argue that an increased violation of local realism is only significative if accompanied by a comparable improvement of the robustness to noise of the test. After all, Bell tests have to be verified in a lab, not on paper.

In a discrete variable setting, the question of the tolerance to noise of a Bell test is often investigated introducing the noise fraction \cite{Krasz}. The noise fraction quantifies the maximum amount of depolarizing noise 
one can add to an entangled state and still detect non-local correlations.
The depolarizing noise is characterized by the state
$\mathbbm{1}/d$, where $d$ is the dimension of the Hilbert space.
However, in the continuous variable regime, the noise model underlying the noise fraction is irrelevant. Even if we deal with a finite number of photons, such as with the truncated photon number correlated states of Section \ref{cpns}, the Hilbert space spanned by the eigenstates of the quadrature operators that are measured remains infinite dimensional. Hence, the appropriate basis is the infinite photon number bases and operators proportional to the identity $\mathbbm{1}$ have no physical meaning. To adopt an objective measure of the magnitude of the violation of local realism, we must therefore introduce a relevant noise model. In this Section, we will consider 
a probabilistic erasure: with probability $p$, the mode of a random party is erased; otherwise it is untouched. This noise acts independently on each mode and transforms an initial state $|\Psi\rangle$ into
\begin{align}\label{rho1m}
\nonumber \rho&=(1-p)^m|\Psi\rangle\langle\Psi| \\
& + p(1-p)^{m-1}\lbrace\sum_{t=1}^m {\mathrm Tr}_t(|\Psi\rangle\langle\Psi|)\otimes|0\rangle_t\langle 0|\rbrace\\
&+ ... +p^m |0\rangle_1\langle 0|\otimes...\otimes|0\rangle_m\langle 0|
\end{align}
Such a probabilistic erasure is known to appear in, e.g. atmospheric transmissions, and has recently been studied in Refs.~\cite{wittmann,niset}.

Let us first consider the photon number correlated states of Section \ref{cpns}, and concentrate on the second term of \refeq{rho1m}. Each element of the sum corresponds to the erasure of one of the subsystems. So, suppose for example that the state of subsystem $m$ has been erased and replaced by vacuum while distributing $|\Psi\rangle$. The corresponding state shared between the $m$ parties reads
\begin{align}
\rho_{1,m}&={\mathrm Tr}_m(|\Psi\rangle\langle\Psi|)\otimes|0\rangle_m\langle 0|\\
\nonumber &=\sum_{r=0}^\infty |c_r|^2|r\rangle_1\langle r |\otimes ... \otimes |r\rangle_{m-1}\langle r |\otimes |0\rangle_m\langle 0 |
\end{align}
This state is diagonal in the photon number bases, hence the results of all possible measurements are equiprobable, i.e., $\forall \theta_1,...,\theta_m$, $P(x(\theta_1),...,x(\theta_m))=cst$, and all correlation coefficients vanish. This was to be expected from photon correlated states as their entanglement is truly $m$-partite; tracing out one subsystem makes the state become separable. Thus, this property also holds for the other noisy terms of (\ref{rho1m}), so that only the erasure-free term $|\Psi\rangle\langle\Psi|$ will contribute to the Bell factor. We obtain
\begin{align} \label{noisyB}
{\cal B}_\rho=(1-p)^m {\cal B}_m
\end{align}

To illustrate this result, consider the $m$-partite GHZ state, \refeq{mGHZ}. The noisy Bell factor reads ${\cal B}_\rho=(1-p)^m \sqrt{2} (\frac{4}{\pi})^{m/2} $, hence the maximum probability of erasure $p_{max}$ such that non-local correlations can be detected is
\begin{align}
p_{max}&=1-\frac{\sqrt{\pi}}{2}2^{1/2m}
\end{align}
This value exponentially tends towards $1-\sqrt{\pi}/2$ as $m$ goes to infinity, and we observe the desired increased robustness to noise as $m$ becomes large.

Finally, consider now the states of Section \ref{mv}. First note that $\langle f|g\rangle=0$, hence in the $\{|f\rangle,|g\rangle\}$ bases, these states look like GHZ states. We thus expect their robustness to noise to behave like the photon correlated number states. Indeed, going through the calculation, one finds that for every noisy term of (\ref{rho1m}), and for every choice of measurement setting the probability is an even function of the results, e.g. if party $m$ looses its mode, the  probability $P(x_1,...,x_m)$ to obtain $x_1,...,x_m$ given that $X_1,...,X_m$ is measured is an even function. As a result, none of the noisy terms contribute to the correlation coefficients, and the noisy Bell factor is again given by (\ref{noisyB}) as expected. 
As a conclusion, with respect to probabilistic erasure, more parties means more robustness.

\section{A candidate 3-mode state}\label{tmsg}
Let us now see how the results of Sec.~\ref{mv} can be exploited in
order to prepare multipartite states such that, on the one hand,
they exhibit a significantly high violation of local realism and, on the
other hand, they may be generated with near future technology. In
particular, we will focus on a class of three-party states whose
generation involves four optical ``Schrodinger-cat'' states,
i.e., four single mode superpositions of coherent states~\cite{NGaus_exp1}.

Before proceeding let us make a remark. The main goal, as said, is now
to exploit the gain expected in the amount of violation of locality
when the number of involved parties is increased. To this aim, a
natural candidate may have been the generalization 
of the photon-subtracted state discussed 
in Refs.~\cite{Patron,Nha}. Unfortunately, in the case of
three parties (and, actually, in the case of any odd number of
parties) the generalization of the strategy adopted in
Refs.~\cite{Patron,Nha} is not effective. This is due to symmetry
reasons and the use of the sign binning. Specifically, on the one
hand, the joint probability distribution
$P(x(\theta_1),...,x(\theta_m))$ of any number of quadratures is an
even function under the exchange of the arguments, on the other hand,
the sign binning introduces an odd function (for odd number of
parties) in the integration that has to be performed in order to
obtain any correlation function. As a consequence, all the correlators
are zero giving trivially no violation.

Let us now see how the increase in the number of parties involved in a
Bell test can be exploited using a different approach, based on the
results given in Sec.~\ref{mv}. In Ref.~\cite{Wenger}, the authors
propose to use a superposition of Gaussians to implement the
functions $f$ and $g$ of \refeq{mv_state}. In particular, in the case of
only two Gaussians they considered the family of states defined by
\begin{subequations}\label{fg3}
\begin{align}
  f(x)&=\bra{x}[c_+(\ket{\alpha}+\ket{-\alpha})] \,,\\
  g(x)&=\bra{x}[c_-(\ket{\alpha}-\ket{-\alpha})] \,,
\end{align}
\end{subequations}
where $c_\pm^2=1/[2(1\pm e^{-2|\alpha|^2})]$.
One can then calculate the corresponding $V$ and $W$ coefficients
using \refeq{VW}, which for large amplitudes $|\alpha|\rightarrow\infty$
gives $V=1$ and $W\simeq0.64$. As noticed in
Ref.~\cite{Wenger}, in the case of two parties no violation is possible, i.e
${\cal B}_2\simeq1.90$. However, as can be seen in \refeq{B3}, one can achieve a violation of ${\cal B}_3\simeq 2.23$ already for three parties.
The corresponding state reads
\begin{equation}\label{ecs3}
  \ket{\Psi_3}=\frac{1}{\sqrt{2}}\left[
c_-^3(\ket{\alpha}-\ket{-\alpha})^{\otimes\,3}+
c_+^3(\ket{\alpha}+\ket{-\alpha})^{\otimes\,3}
\right]\,.
\end{equation}
where we have put $\theta=0$.

Since the maximum violation is achieved for large amplitudes, we can
 consider the following simpler state:
\begin{multline}\label{3ps}
  \ket{\Psi'_3}=c'(\ket{\alpha, \alpha, \alpha}+
\ket{\alpha, -\alpha, -\alpha}+ \\
\ket{-\alpha, \alpha, -\alpha}+\ket{-\alpha, -\alpha, \alpha})\,,
\end{multline}
which coincides with $\ket{\Psi_3}$ for $|\alpha|\rightarrow\infty$
[where we defined $c'^2=1/[4(1+3e^{-4 |\alpha|^2})$].  In order to
obtain the Bell factor ${\cal B}_3$ corresponding to such a state, we
calculated the probabilities of the binned outcomes using \refeq{fg3}
to define the roots, in combination with Eqs.~(\ref{prob}) and (\ref{pquad}).
Specifically, we considered the domains $D_{x}^\pm$ and $D_{p}^\pm$
inherited by the state $\ket{\Psi_3}$:
\begin{subequations}\label{domains3}
\begin{align}
  D_{x}^+&=\{x\in\R|x\geq 0\}\\
  D_{x}^-&=\{x\in\R|x < 0\}\\
  D_{p}^+&=\{p\in\R|-\cos(p\alpha)\sin(p\alpha)\geq 0\}\\
  D_{p}^-&=\{p\in\R|-\cos(p\alpha)\sin(p\alpha) < 0\}\;.
\end{align}
\end{subequations}

\begin{figure}
  {\includegraphics[width=0.45\textwidth]{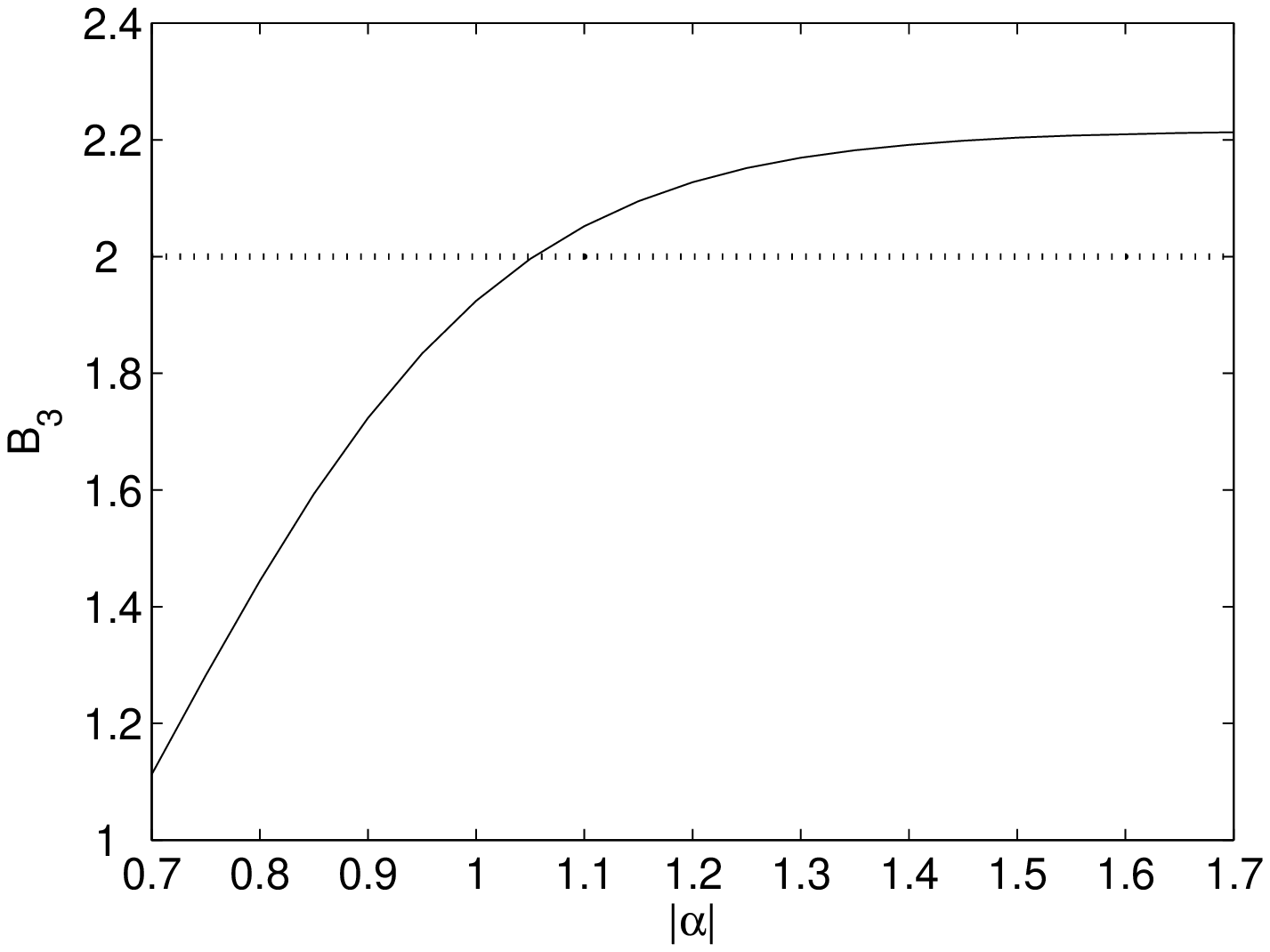}}
  \caption{Bell factor ${\cal B}_3$ for the state $\ket{\Psi'_3}$ as a
    function of the amplitude $\alpha$ [see \refeq{3ps}]. }
\label{psit_b3}
\end{figure}

The Bell coefficient ${\cal B}_3$ calculated with such a procedure is
shown in Fig.~\ref{psit_b3}.
One can see that for amplitudes as small as $|\alpha|\simeq1.1$,
the state $\ket{\Psi'_3}$ already gives values above the local bound.
We note that in this regime of small amplitudes, $\ket{\Psi'_3}\neq\ket{\Psi_3}$
and the domains defined in Eqs.~(\ref{domains3}) might be non-optimal. As $\alpha$ is
increased, a violation around $10\%$ of the MK inequality is rapidly achieved.

\begin{figure}
  {\includegraphics[width=0.45\textwidth]{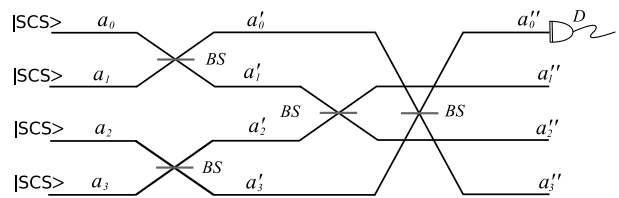}}
  \caption{Schematic of a possible way to conditionally generate the
    state $\ket{\Psi'_3}$ in \refeq{3ps}: ($\ket{\rm SCS}$)
    superposition of coherent states [see \refeq{scs}]; (BS) balanced
    beam splitter; (D) homodyne detector.}
\label{psit_gen}
\end{figure}

Now let us describe how the state $\ket{\Psi'_3}$ may be conditionally
generated by using linear optics and superpositions of coherent states (SCS)
of this form:
\begin{equation}\label{scs}
  \ket{\rm SCS}=c_+(\ket{\alpha}+\ket{-\alpha}).
\end{equation}
Consider the scheme depicted in Fig.~\ref{psit_gen}.
Two copies of the state $\ket{\rm SCS}$ in mode $a_0$ and $a_1$ are
mixed in a balanced beam splitter. The same action is performed on
modes $a_2$ and $a_3$. Then modes $a'_1$ and $a'_2$, as well as $a'_0$
and $a'_3$ are respectively mixed by means of two other beam
splitters.  As a last step mode $a''_0$ is measured via a homodyne
detector. It is straightforward to show that, when the measurement
outcome $-\alpha$ is obtained, then the conditional state of the
remaining modes coincides approximately with $\ket{\Psi'_3}$.

The main issue in implementing the scheme above concerns the
generation of the four states $\ket{\rm SCS}$, whose preparation is
experimentally demanding.  However, their generation in traveling
light modes has been recently reported by several groups
\cite{NGaus_exp1}. Thus, one may envisage that, in the future,
the implementation of the whole scheme of Fig.~\ref{psit_gen} will be
possible.

\section{Conclusions}\label{esco}

In this paper, we have investigated the violation of local realism in Bell tests
based on homodyne measurements performed on multipartite continuous-variable
states. We have proven that the Mermin-Klyshko inequality supplemented with sign binning can be violated by an amount which grows exponentially with the number of parties. Furthermore, we have shown that it is possible to attain the maximal violation allowed by quantum mechanics by tuning the state and the binning strategy appropriately.
The benefit of this multipartite approach was then shown to be effective 
in practice
by analyzing the increased robustness to noise of the Bell tests with the
number of modes.

It is worthwhile noting that our results are not based on a direct mapping 
from continuous to discrete variables. Such a procedure would in fact force  a particular measurement to be used in the locality test, as the analysis of Ref.~\cite{pseudospin} shows. We instead consider homodyne measurement because of its high detection efficiency. In this case, the possibility to
obtain a maximal violation of the Mermin-Klyshko inequality is 
non-trivial, since
(i) homodyne measurement represents a small subset 
of all possible measurements and (ii) the
binning procedure may cause an irreversible loss of information. As it
turns out, however, such a loss is not detrimental
if suitable binning procedures are used,
properly adapted to the states under investigation.

The experimental significance of our findings lies in that
one can effectively benefit from the increased violation of locality 
in the considered multipartite continuous-variable scenario. 
In general, this allows for a greater freedom in the search for 
feasible multipartite states that can be used to test nonlocality 
in an actual experiment suffering from noise. As an illustration, 
inspired by the analysis of the states that
maximally violate the local bound, we have proposed an approximation of the
latter state which may be experimentally realized in a reasonable future. This state involves three parties and gives a violation 
of around $10\%$ of the local bound.
As a perspective of this work, it is expected that other states of light with similarly high violation and suitably adapted to experimental constraints might be found, inspired by our results. 

\section{Acknowledgments}

We gratefully acknowledge the financial support from the EU Projects QAP and COMPAS. A.A. and A.F. acknowledge the Spanish MEC projects FIS2007-60182 and Consolider Ingenio 2010 QOIT, and the Juan de la Cierva grant. N.J.C. and J.N. acknowledge financial support from the IUAP programme of the Belgian government under project Photonics@be. Finally, J.N. acknowledges
support from the Belgian FRIA foundation.



\end{document}